\documentclass[twocolumn,twoside,letterpaper,11pt]{article}
\usepackage{graphicx,xrrc,natbib}
\usepackage{times}

\def\ltsim{\raise 2pt \hbox {$<$} \kern-1.1em \lower 4pt \hbox {$\sim$}}
\def\gtsim{\raise 2pt \hbox {$>$} \kern-1.1em \lower 4pt \hbox {$\sim$}}
\begin{document}

% The title is completely in capital letters and has no ending period

\title{THE COOLING FLOW CLUSTER ABELL 2626 AND THE ASSOCIATED RADIO EMISSION}

\author{    Myriam Gitti                       } % author(s)
\institute{ Institute of Astrophysics, University of Innsbruck} % the institute
\address{   Technikerstrasse 25, A-6020 Innsbruck, Austria  } % street address
\email{     myriam.gitti@uibk.ac.at                         } % email addresses

\maketitle

%%%%%%%%%%%%%%%%%%%%%%%%%%%%%%%%%%%%%%%%%%%%%%%%%%%%%%%%%%%%%%%%%%%%%%%%%%%%%%

\abstract{ 
We present VLA data at 330 MHz and 1.5 GHz of the radio emission observed 
in the cooling flow cluster A2626.
By producing images at different resolutions we found that the radio source 
consists of different components: an unresolved core plus a jet-like feature, 
two elongated parallel features, and an extended diffuse emission 
(radio mini-halo).
Low resolution images allow us to derive morphological and spectral 
information of the diffuse emission: the radio mini-halo is extended on a 
scale comparable to that of the cooling flow region and is characterized 
by amorphous morphology, lack of polarized flux and very steep spectrum which
steepens with distance from the center.
We then applied to this new mini-halo source a model for particle 
re--acceleration in cooling flows (Gitti, Brunetti \& Setti 2002).
In particular, we found that its main radio properties (brightness profile, 
integrated radio spectrum and radial spectral steepening) can be accounted 
for by the synchrotron radiation from relic relativistic electrons in the 
cluster, which are efficiently re-accelerated by MHD turbulence amplified 
by the compression of the cluster magnetic field in the cooling flow region.}

%%%%%%%%%%%%%%%%%%%%%%%%%%%%%%%%%%%%%%%%%%%%%%%%%%%%%%%%%%%%%%%%%%%%%%%%%%%%%

\section{Introduction}

The cluster A2626 (z=0.0604) is a good candidate to study the interaction 
between the X--ray emitting intra--cluster medium (ICM) and radio emitting 
plasma in clusters of galaxies.  
This cluster hosts a relatively strong cooling flow (White, Jones \& Forman 
1997) and contains a very unusual radio source exhibiting a compact unresolved
core and a diffuse structure (Roland et al. 1985; Burns 1990). Earlier radio 
observations showed that the compact component is associated with the centrally
dominant elliptical galaxy (Owen, Ledlow, \& Keel 1995), while the diffuse
emission has no optical counterpart. Comparisons with X--ray data revealed an 
enhanced X--ray emission spatially coincident with the radio source,
thus providing strong observational evidence for a connection between the hot,
X--ray gas and the radio plasma. In addition, the X--ray map of the cooling 
flow region shows an elongation coincident with the diffuse radio component 
(Rizza et al. 2000). 

We studied in detail the radio properties of A2626 in order to better 
investigate the nature of the interaction between the ICM and the radio plasma
and in particular the origin of the diffuse radio emission in the core of 
cooling flow clusters. 

A Hubble constant
$\mbox{H}_0 = 50 \mbox{ km s}^{-1} \mbox{ Mpc}^{-1}$ 
is assumed in this paper, therefore at the distance of A2626  
$1'$ corresponds to $\sim$ 95 kpc.
The radio spectral index $\alpha$ is defined such as 
$S_{\nu} \propto \nu^{-\alpha}$.

%%%%%%%%%%%%%%%%%%%%%%%%%%%%%%%%%%%%%%%%%%%%%%%%%%%%%%%%%%%%%%%%%%%%%%%%%%%%%%%

\section{Radio properties of A2626: analysis of VLA archive data}

We have analyzed VLA archive data of A2626 at the frequencies $\nu = 330$ MHz
and $\nu=1.5$ GHz and at different resolutions with the aim to derive the 
surface brightness map, the total spectral index and the spectral index 
distribution of the diffuse radio emission.
Standard data reduction was done using the National Radio Astronomy Observatory
(NRAO) AIPS package.

%%%%%%%%%%%%%%%%%%%%%%%%%%%%%%%%%%%%%%%%%%%%%%%%%%%%%%%%%%%%%%%%%%%%%%%%%%%%%%%

\subsection{Radio analysis at $\nu=1.5$ GHz}

\begin{figure}
\includegraphics{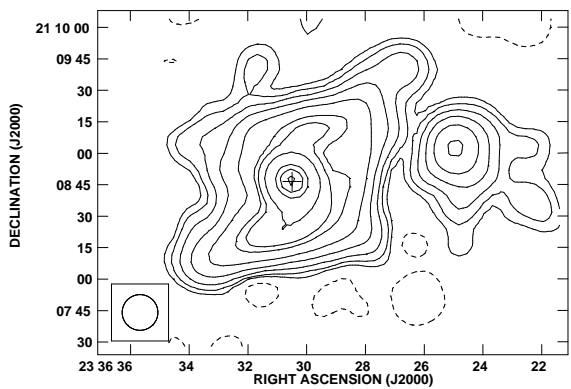}
\vspace{5cm}
\caption{\small
1.5 GHz VLA map of A2626 at a resolution of $17'' \times 17''$. 
The contour levels are $-0.06$ (dashed), 0.06, 0.12, 0.24, 0.48, 0.96, 1.92,
2.5, 4, 8, 10, 12 mJy/beam. The r.m.s. noise is 0.02 mJy/beam.
The cross indicates the position of the cluster center (Gitti et al. 
2004).}
\label{fig:8.12_gitti_fig1}
\end{figure}

We used the 1.5 GHz C array data to produce low resolution image with a 
circular restoring beam of 17 arcsec. 
The 1.5 GHz map (Fig. \ref{fig:8.12_gitti_fig1}) shows an unresolved core and 
a diffuse boxy--shaped emission extended for $\sim 2'$, 
corresponding to about 190 kpc.
No significant polarized flux is detected, leading to a 
polarization upper limit of $<2$ \%. 
An unrelated source (associated with the S0 galaxy IC5337 at the same 
redshift of the cluster) is present to the west of the diffuse emission, 
with a total flux density of $\sim$3.9 mJy.

\begin{figure}
\includegraphics{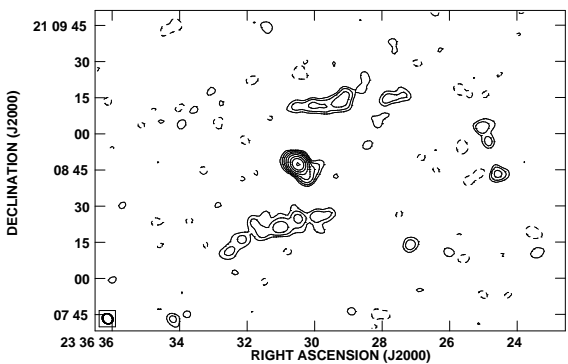}
\vspace{5cm}
\caption{\small
1.5 GHz VLA map of A2626 at a resolution of $4.5'' \times 3.9''$. 
The contour levels are $-0.09$ (dashed), 0.09, 0.18, 0.36, 0.7, 1.4, 3,
5.5, 11, 20 mJy/beam. The r.m.s. noise is 0.03 mJy/beam (Gitti et al. 2004).}
\label{fig:8.12_gitti_fig2}
\end{figure}

\begin{figure}
\includegraphics{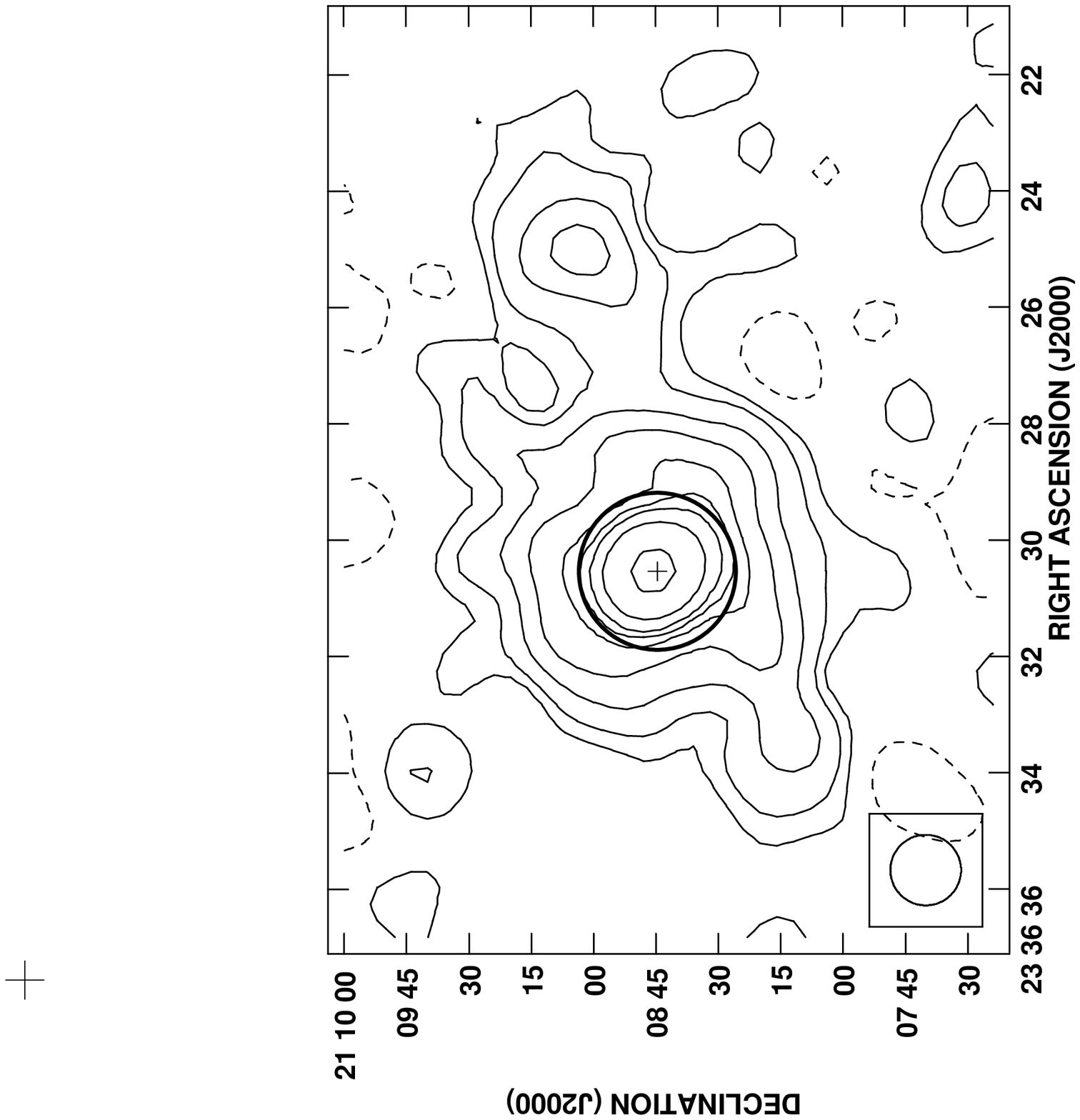}
\vspace{5cm}
\caption{\small
1.5 GHz VLA map of A2626 at a resolution of $17'' \times 17''$
after the subtraction of the two elongated features visible in
Fig. \ref{fig:8.12_gitti_fig2}. The cross indicates the position of the cluster
center. 
The contour levels are $-0.18$ (dashed), 0.18, 0.36, 0.72, 1.2, 1.85, 2.3
mJy/beam. The r.m.s. noise is 0.08 mJy/beam.
The circle defines the nuclear region excluded in modeling the 
diffuse radio emission.
}
\label{fig:8.12_gitti_fig4}
\end{figure}

By using the 1.5 GHz B array data we produced the high resolution
image (Fig. \ref{fig:8.12_gitti_fig2}) with a restoring beam of 
$4.5 \times 3.9$ arcsec.
The central component is found to consist of an unresolved
core, plus a jet-like feature pointing to the south-western direction.
The extended emission is resolved out and two elongated parallel features of 
similar brightness and extent are detected. 
The flux density of the central component, the short jet included,
is $\simeq 13.5$ mJy, in agreement with Roland et al. (1985).

In order to estimate the total integrated flux density
of the diffuse radio emission at 1.5 GHz and derive the surface 
brightness map and spectral trend, we subtracted the emission from the central 
nuclear source.
This subtraction allows us to derive a good estimate of the 
total flux density of the diffuse emission ($S_{1.5} \sim 29$ mJy).

As discussed in Gitti et al. (2004), 
we believe that the elongated structures visible in Fig. 
\ref{fig:8.12_gitti_fig2} are distinct 
(or that they may represent an earlier evolutionary stage)
from the diffuse emission, and that the diffuse radio source
may belong to the mini--halo class.
The total flux density of these structures is $\sim 6.6$ mJy, 
thus contributing to only $\sim$ 20 \% of the flux of 
the diffuse radio emission.
Since we are interested in studying and modeling the diffuse radio emission,
we produced a new low--resolution map at 1.5 GHz where these discrete
radio features have been subtracted (Fig. \ref{fig:8.12_gitti_fig4}).
To be conservative, the region in which the surface brightness 
is affected by the central emission (within the circle 
in Fig. \ref{fig:8.12_gitti_fig4}) has been excluded in modeling the diffuse 
radio emission (see Sect. \ref{8.12_gitti_model.sec}).

%%%%%%%%%%%%%%%%%%%%%%%%%%%%%%%%%%%%%%%%%%%%%%%%%%%%%%%%%%%%%%%%%%%%%%%%%%%%%%%

\subsection{Radio map at $\nu=330$ MHz}

\begin{figure}
\includegraphics{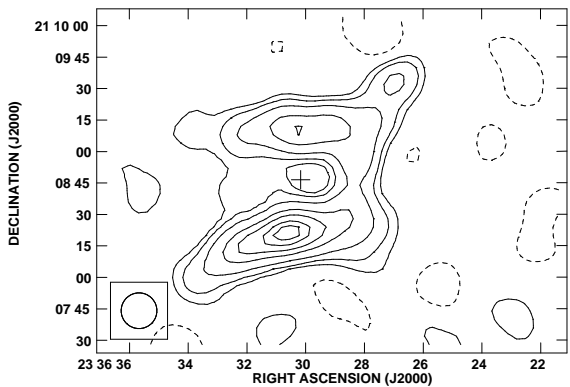}
\vspace{6cm}
\caption{\small
330 MHz VLA map of A2626 at a resolution of $17'' \times 17''$. 
The contour levels are $-8.5$ (dashed), 8.5, 17, 34, 68, 100, 135,
160 mJy/beam. The r.m.s. noise is 3.1 mJy/beam. The cross indicates the 
position of the cluster center.}
\label{fig:8.12_gitti_fig3}
\end{figure}

We used the 330 MHz B+DnC array data to produce low resolution images with a 
circular restoring beam of 17 arcsec. The imaging procedure was performed using
data with uv coverage matching that at 1.5 GHz. 
The diffuse structure at 330 MHz (Fig. \ref{fig:8.12_gitti_fig3}) consists of
two elongated almost parallel features located
to the north and south of the core, respectively.
No radio emission is detected at the location of the
1.5 GHz radio nucleus. Assuming for the 330 MHz core flux
an upper limit of 3$\sigma$, we obtain that 
the radio core  has an inverted spectrum (Table \ref{8.12_gitti_risradio.tab}).
The total flux density of the diffuse emission is $S_{330} \sim 1$ Jy. 
The unrelated source, detected at 1.5 GHz to the west of the diffuse
radio emission, is not detected at 330 MHz; this implies
a spectrum with $\alpha^{1.5}_{0.3}$ \ltsim 0.6.

%%%%%%%%%%%%%%%%%%%%%%%%%%%%%%%%%%%%%%%%%%%%%%%%%%%%%%%%%%%%%%%%%%%%%%%%%%%%%%%

\subsection{Spectral index distribution}

Due to the lack of a high--resolution image,
the two elongated features cannot be subtracted at 330 MHz as done at 1.5
GHz. Therefore, in deriving the spectral information 
of the diffuse emission we considered for consistency
the low--resolution images in Fig. \ref{fig:8.12_gitti_fig1} 
and \ref{fig:8.12_gitti_fig3}, which both include the 
contribution of the two features to the total flux.
The integrated spectral index of the diffuse emission between 
$\nu=330$ MHz and $\nu = 1.5$ GHz is $\alpha \sim 2.4$. 

\begin{figure}
\includegraphics{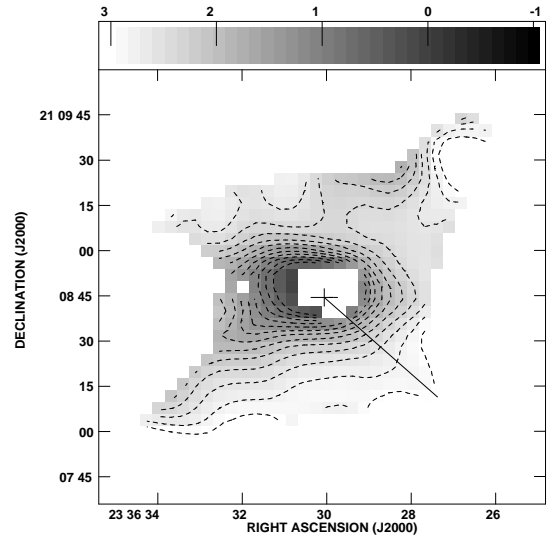}
\vspace{7cm}
\caption{\small
Spectral index distribution between $\nu=330$ MHz and $\nu = 1.5$ GHz
at a resolution of $17'' \times 17''$. The contours are in steps of 0.2, 
where the lightest grey is in the range 2.8 to 3.
The lighter the grey, the steeper the spectral index.
We have excluded the region in which the error is $>0.2$.
The cross indicates the position of the subtracted radio core and the solid 
line represent the direction we have considered for radial profiles (Gitti et 
al. 2004).}
\label{fig:8.12_gitti_fig5}
\end{figure}

In Fig. \ref{fig:8.12_gitti_fig5} we show a grey scale image
of the spectral index map of the diffuse radio emission between  
$\nu=330$ MHz and $\nu = 1.5$ GHz.
The spectrum steepens from the central region towards the north and south
direction, with the spectral index increasing from $\sim$1.2
to $\sim$3. 

%%%%%%%%%%%%%%%%%%%%%%%%%%%%%%%%%%%%%%%%%%%%%%%%%%%%%%%%%%%%%%%%%%%%%%%%%%%%%%%

\subsection{Summary}

The radio results are summarized in Table \ref{8.12_gitti_risradio.tab}.

\begin{table}
\begin{center}
\small
\caption{\small Radio results for A2626}
\begin{tabular}{cccc}
\hline
\hline
Emission & $S_{330}$ & $S_{1.5}$ & $\alpha$ \\
~&  (mJy) & (mJy) & ($S_{\nu} \propto \nu^{-\alpha}$)\\   
\hline
~&~&~\\
Nuclear & $<$9.3 & $13.5 \pm 1.5$ & \ltsim $-0.25$\\
Diffuse & $990 \pm 50$ & $29 \pm 2$ & $2.37 \pm 0.05$ \\
\hline
\label{8.12_gitti_risradio.tab}
\end{tabular}
\end{center}
\end{table}

The extended radio source observed at the center of A2626 is characterized by 
amorphous morphology, lack of polarized flux and a very steep spectrum that
steepens with distance from the center.
Finally, the size of the diffuse radio emission is comparable 
to that of the cooling flow region (Rizza et al. 2000). 
These results indicate that the diffuse radio source may be 
classified as a mini--halo.

%%%%%%%%%%%%%%%%%%%%%%%%%%%%%%%%%%%%%%%%%%%%%%%%%%%%%%%%%%%%%%%%%%%%%%%%%%%%%%%

\section{Model for electron re--acceleration in cooling flows: origin of radio 
mini--halos}
\label{8.12_gitti_model.sec}

%%%%%%%%%%%%%%%%%%%%%%%%%%%%%%%%%%%%%%%%%%%%%%%%%%%%%%%%%%%%%%%%%%%%%%%%%%%%%%%

\subsection{Overview of the model}

Gitti, Brunetti \& Setti (2002) developed a model for radio mini--halos
consisting in the re--acceleration of relic relativistic electrons
by MHD turbulence \textit{via} Fermi--like processes, with the necessary 
energetics supplied by the cooling flow. 

\begin{figure}
\includegraphics{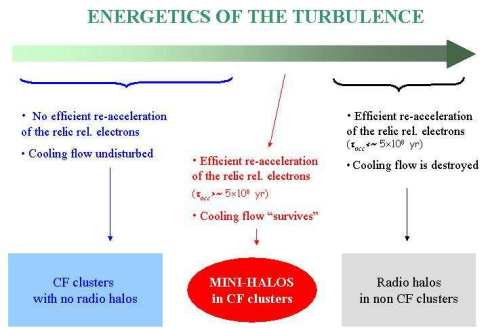}
\vspace{5cm}
\caption{\small
General picture of the model
}
\label{fig:8.12_gitti_fig6}
\end{figure}

The MHD turbulence is assumed to be frozen into the flow of the thermal ICM 
and is thus amplified due to the compression of the turbulent coherence length
scale and the amplification of the magnetic field in the cooling flow
region. However, the seed large--scale turbulence (before the amplification) 
must not be too high in order to avoid the disruption of the cooling flow. 
There is thus a fine--tuning of the turbulent energy to produce diffuse radio 
emission from a cooling flow cluster: in the picture of the model the 
physical conditions of mini--halo clusters are intermediate between
those which lead to the formation of extended radio halos, hosted by 
non--cooling flow clusters, and those holding in cooling flow clusters without 
radio halos (see Fig. \ref{fig:8.12_gitti_fig6}).

%\begin{figure}
%  \begin{center}
%    \includegraphics[width=\columnwidth]{8.12_gitti_fig6.ps}
%    \caption{\small General picture of the model}
%    \label{fig:8.12_gitti_fig6}
%  \end{center}
%\end{figure}

%%%%%%%%%%%%%%%%%%%%%%%%%%%%%%%%%%%%%%%%%%%%%%%%%%%%%%%%%%%%%%%%%%%%%%%%%%%%%%%

\subsection{Model application to A2626}

In order to test the predictions of the model we have calculated the following 
expected observable properties:
total spectrum, brightness profile, and steepening of the spectrum with the 
distance from the center.
The expected brightness profile and radial spectral steepening 
in the model were compared to the observed profiles along the radial 
direction indicated in Fig. \ref{fig:8.12_gitti_fig5}.
We found that the model is able to reproduce all the observational constraints 
of A2626 for physically--meaningful values of the parameters 
(Table \ref{8.12_gitti_model.tab}). The application of the model to A2626
is discussed in detail in Gitti et al. (2004). 

\begin{table}
\begin{center}
\small
\caption{\small Model results for A2626}
\begin{tabular}{ccc}
\hline
\hline
Magnetic field & Electron & $P_{\rm reacc}/P_{\rm CF}$\\
at the cooling radius & energetics & ~ \\   
\hline
~&~&~\\
1.2-1.6 $\mu$G & $2.2 \times 10^{57}$ erg & 0.7 \%\\
\hline
\label{8.12_gitti_model.tab}
\end{tabular}
\end{center}
\end{table}

%%%%%%%%%%%%%%%%%%%%%%%%%%%%%%%%%%%%%%%%%%%%%%%%%%%%%%%%%%%%%%%%%%%%%%%%%%%%%%%

\subsection{Conclusions}

The main radio properties of the mini--halo observed in A2626 can be accounted
for by the model. We found that only a small fraction of the cooling flow power
should be converted into electron re--acceleration. Therefore we conclude
the cooling flow process is able to provide sufficient energy to power the
radio mini--halo in A2626. 

%%%%%%%%%%%%%%%%%%%%%%%%%%%%%%%%%%%%%%%%%%%%%%%%%%%%%%%%%%%%%%%%%%%%%%%%%%%%%%%

\section*{Acknowledgments}

It is a pleasure to acknowledge my collaborators G. Brunetti, L. Feretti 
and G. Setti. I would also like to thank K. Dyer and L. Sjouwerman 
for organizing this intresting Conference.
This work was partly supported 
by the CNR grant CNRG00CF0A, 
by the Italian 
Ministry for University and Research (MIUR) under grant Cofin 2001-02-8773 and 
by the Austrian Science Foundation FWF under grant P15868.


\begin{thebibliography}{}
\setlength\itemsep{0cm}

\bibitem[Burns (1990)]{burns} Burns J. O. 1990, AJ 99, 14

\bibitem[Gitti, Brunetti \& Setti (2002)]{gitti02} Gitti, M., Brunetti, G., \& Setti, G. 2002, A\&A, 
             386, 456 

\bibitem[Gitti et al. (2004)]{gitti04} Gitti, M., Brunetti, G., Feretti, L. \& Setti, G. 2004, A\&A, 
             417, 1

\bibitem[Owen, Ledlow \& Keel (1995)]{owen} Owen, F.N., Ledlow, M.J., Keel, W.C. 1995, AJ, 109, 140

\bibitem[Rizza et al. (2000)]{rizza} Rizza E., Loken C., Bliton M., Roettiger K., \& Burns J.O. 
             2000, AJ, 119, 21

\bibitem[Roland et al. (1985)]{roland} Roland, J., Hanish, R. J., V\'eron P., \& Fomalont, E. 1985,
             A\&A, 148, 323

\bibitem[White, Jones \& Forman (1997)]{white} White, D. A., Jones, C., \& Forman, W. 1997, MNRAS, 292, 419

\end{thebibliography}
\end{document}